בס"ד

# Monte Carlo Simulations for Ghost Imaging Based on Scattered Photons

Shukrun R. H [1,2]., Klein Y[1]., Sefi O[1]., Fried Y[2]., Epstein L[2]., Shwartz S*[1].

[1]*Physics Department and Institute of Nanotechnology, Bar-Ilan University, 52900 Ramat Gan, Israel.*
[2] *Radiation Safety Department, Soreq Nuclear Research Center, Yavne 81800, Israel*

* Corresponding author's e-mail: sharon.shwartz@biu.ac.il

**Abstract.** X-ray based imaging modalities are widely used in research, industry, and in the medical field. Consequently, there is a strong motivation to improve their performances with respect to resolution, dose, and contrast. Ghost imaging (GI) is an imaging technique in which the images are reconstructed from measurements with a single-pixel detector using correlation between the detected intensities and the intensity structures of the input beam. The method that has been recently extended to X-rays provides intriguing possibilities to overcome several fundamental challenges of X-ray imaging. However, understanding the potential of the method and designing X-ray GI systems pose challenges since in addition to geometric optic effects, radiation-matter interactions must be considered. Such considerations are fundamentally more complex than those at longer wavelengths as relativistic effects such as Compton scattering become significant. In this work we present a new method for designing and implementing GI systems using the particle transport code FLUKA, that rely on Monte Carlo (MC) sampling. This new approach enables comprehensive consideration of the radiation-matter interactions, facilitating successful planning of complex GI systems. As an example of an advanced imaging system, we simulate a high-resolution scattered photons GI technique.

## 1 INTRODUCTION

Conventional X-ray imaging modalities measure the transmission through an object using pixelated detectors. The beam that irradiates the sample is nearly uniform in the irradiated area and the image is reconstructed from the non-uniform intensity profile at the detector, which represents the variation between the transmission of different locations of the sample. In the high energy photons, far above the resonance energies, this image of transmission is proportional mainly to the variation in the electron density. Thus, X-ray imaging can be used to infer the structure and even the chemical composition of the inspected object [1]. The image resolution of those modalities is limited by the resolution of the pixelated detector, which is typically in the range of several hundred microns. The dose and contrast are limited by detector efficiency and various noise sources such as scattering, which is a major factor in the photon energy range that is used for medical and industrial imaging.

Ghost imaging (GI) is an imaging technique that was demonstrated with entangled photons in the context of nonlocal quantum correlations [2]. Later it has been realized that two classical beams with correlated intensity distributions can be used for GI and the difference between the quantum and classical GI has been extensively investigated [3–6]. Computational GI (CGI) is a form of classical GI, in which one of the beams is substituted by a computer-controlled mask [7]. One advantage of CGI is that it requires only one single-pixel detector for the measurement of the object and its introduction has led to the development of compressive GI [8]. More recently machine learning approaches have been applied for CGI [9], enabling significant reductions in the measurement time and the dose, as well as improvements in the resolution and contrast of the image [8, 10]. In addition, it has been shown that CGI is more robust to scattering than conventional imaging methods [11, 12].

The successful demonstrations of GI at X-ray wavelengths [13–20] in recent years have opened new possibilities for leveraging the advantages of GI for x-ray imaging. However, the design of GI systems has predominantly relied on relatively simple calculations that solely consider the absorption of the object. This approach overlooks the complex radiation-matter interaction and its implications on GI, particularly the significant influence of photon scattering at high photon energies. Accounting for these intricate interactions presents a formidable challenge in the design of high-photon energy GI.

One potential approach to address this challenge is to use Monte Carlo (MC) sampling codes. These codes rely on random walk techniques to solve the radiation transport problem by repeatedly tracking a single primary particle and recording the interactions along its path [21]. Subsequently, these recorded data are combined to obtain measurable physical quantities. Higher statistical accuracy can be achieved by increasing the number of initial primary particles. MC codes are considered the gold standard for obtaining accurate and reliable solutions for various physical calculations, especially in complex geometries or radiation sources. Moreover, with the increasing availability of computational resources, MC radiation transport codes and programs are now becoming more and more widespread [22]. However, despite the demonstrated successes of MC programs, their adaptation for simulating GI systems has not been reported thus far.

In this work we propose and demonstrate an MC-based approach for analyzing and designing GI systems using the MC particle transport code FLUKA [23, 24] and the Flair [25] graphical user interface. The FLUKA code is widely used in various nuclear and particle physics applications, such as accelerator design, radiation protection and shielding calculations, particle physics research, medical imaging, and radiotherapy. Our approach allows us to simulate GI for complex beams and objects and to provide a detailed insight on the radiation-matter interactions supporting the imaging process. As an illustration, we demonstrate a simulation of a novel approach for scattering-based GI.

## 2  METHODS

There are several variations of GI systems, here we consider a system comprising a radiation source, a mask, an object, and a single-pixel detector. Figure 1 illustrates the correspondence between the laboratory setup and the MC implementation. To incorporate our GI system into the MC code, we need to consider three elements: the source beam, the geometries of the objects (physical dimensions and materials), and the detectors as we elaborate below.

בס"ד

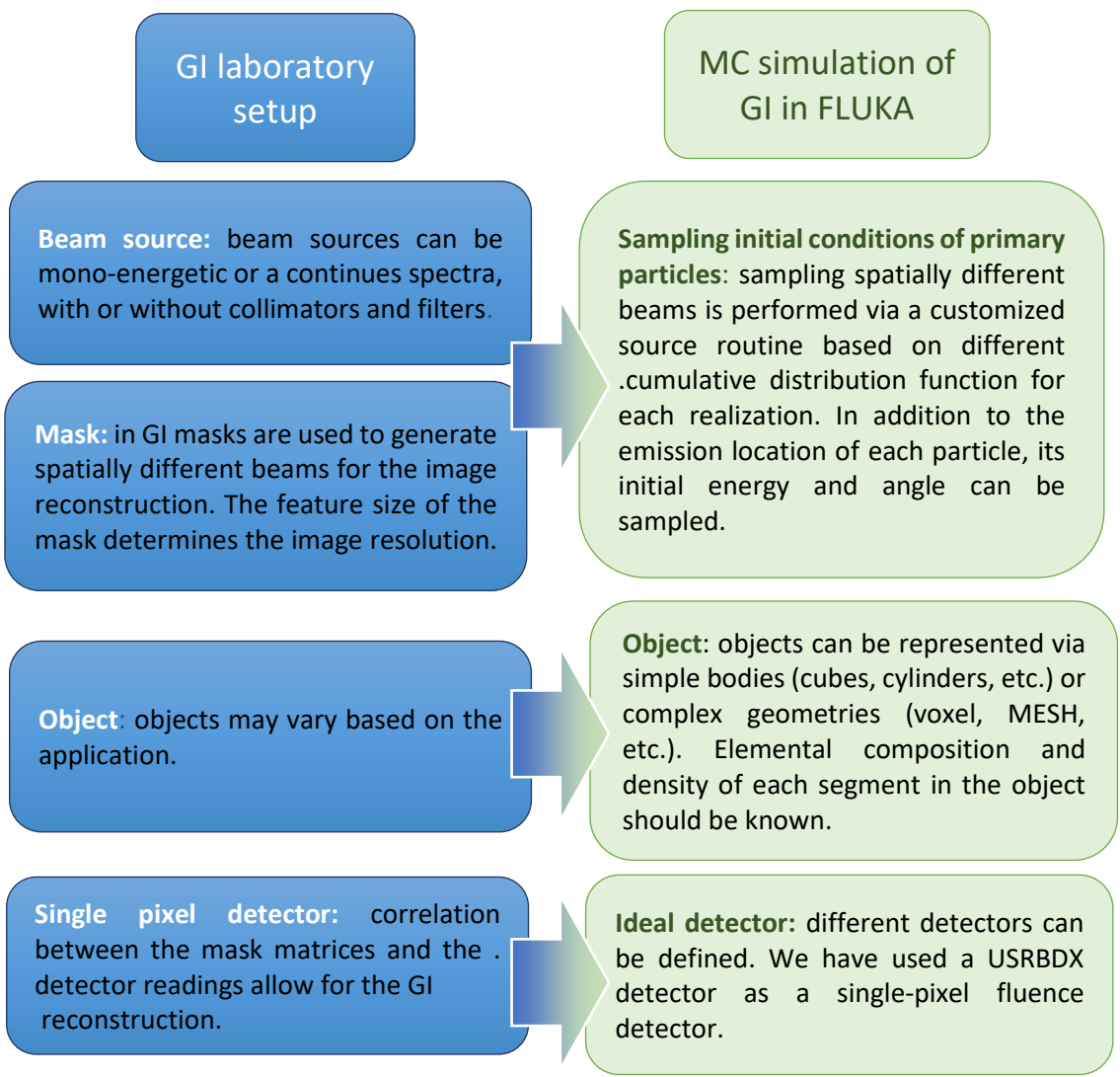

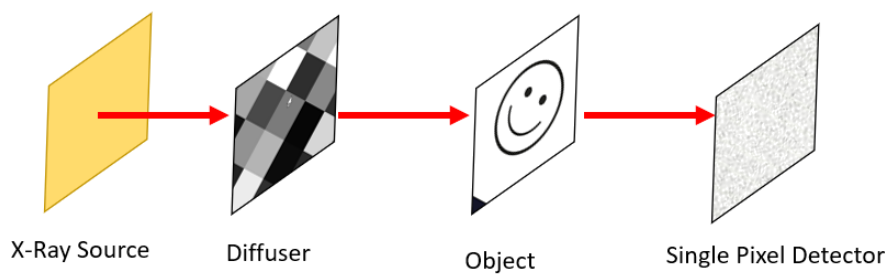

**Figure 1: Ghost imaging setup in the laboratory and the corresponding setup in Monte-Carlo simulations**

## *2.1 Source beam and masks*

Spatially modulated beams are created in the laboratory using masks that modulate the emission patterns of the beam. In GI, the resolution of the reconstructed image depends on the properties of the mask, thus, it is critical to define the mask appropriately. The characteristics of the beam are determined by the laboratory equipment. In the case of X-ray GI, the source setup typically includes an X-ray tube, filtrations, and collimators. Masks can be made from heavy metals such as silver to provide high emission-transmission contrast.

Defining an MC simulation setup starts by determining the probability density functions (PDFs) of each initial condition of the primary particle. However, directly sampling from the PDF can be challenging since computers usually sample random variables uniformly over the interval (0,1). Consequently, we use the cumulative distribution function (CDF) that represents the probability of choosing a variable that is smaller than a specific value x. Since we consider probabilities, the CDF values are within the interval (0,1), similar to the probability space of the numbers that we can generate. Now we can sample from any desired PDF; we generate a uniformly distributed variable, consider it a CDF value, and find the correlated initial condition value.

In MC simulations, each physical object must be modeled. One approach is to model each part of the source setup (X-ray vacuum tube, filtrations, etc.) and the masks. However, this approach is very computational demanding and requires detailed information about specific instruments. Consequently, the more common approach is to sample the initial conditions of the primary particle as it exits the source setup, and only model the object geometry, as described in the following section.

For the simulations we used the MC code FLUKA 4-3.1 [23, 24] and the Flair 3.2-2 [25] graphic user interface. In FLUKA 8 initial conditions must be specified for each primary particle: particle type, energy (E), starting point (x,y,z) and angle cosines in regard to the coordinate system (cos_x, cos_y, cos_z). Each initial condition (except particle type) can be sampled directly using the default BEAM and BEAMPOS cards, but only when sampling from common distributions such as Gaussian or isotropic distributions. More complex beams can be implemented using user defined source routines.

In this work we used a source routine to sample the spatially inhomogeneous beam immediately after the mask. We considered a monochromatic beam at 200 keV that propagates in parallel to the optical axis. The initial location was sampled using random binary masks that were created with Matlab, representing the varying emission-transmission patterns. The masks were normalized and converted to cumulative distribution functions (CDF) to allow sampling in MC simulation. Figure 2 presents examples of emission probability distribution and cumulative probability of the mask. The emission probability distribution represents the transmission pattern of the mask, as we need to simulate the beam immediately after the mask, as explained previously.

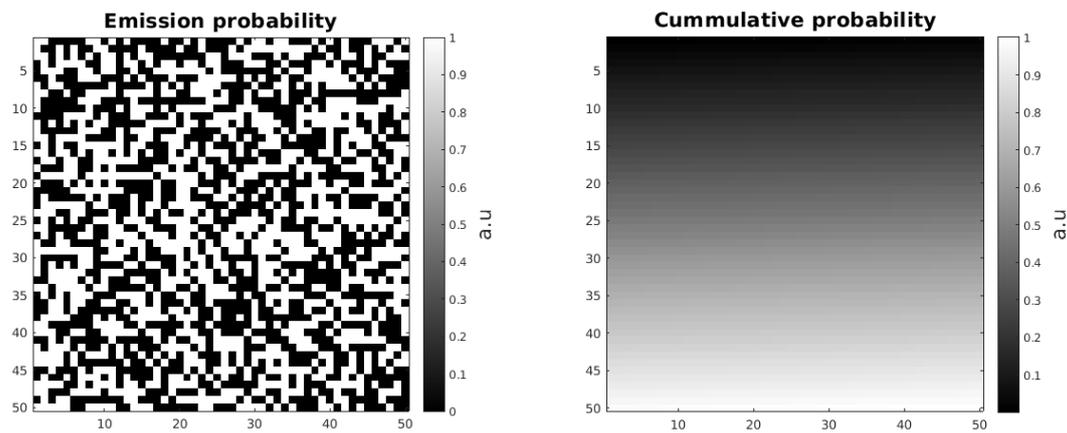

**Figure 2: Emission probability and cumulative probability masks**

## *2.2  System geometries*

The object was defined as two rectangular plates made of iron, located 5 mm from the X-ray source. The imaging system was set in air. Dimensions and material descriptions of each element in the imaging system are presented in table 1.

**Table 1: simulation geometry and materials**

|  |  | Dimensions (h·w·d) (mm³) | material | Density (g/cm³) | Elemental composition (% by mass) |
|---|---|---|---|---|---|
| **Object** | **Big plate** | 3·3·0.5 | Iron | 7.874 | Fe-100 |
|  | **Small plate** | 2·2·0.5 |  |  |  |
| **Surrounding media** |  | -- | Air | $1.20484 \cdot 10^{-3}$ | C-1.248, N-75.5267, O-23.1781, Ar-1.2827 |

## *2.3  Detectors*

In MC simulations it is possible to define various types of detectors, such as fluence, current, energy deposition, and particle track length detectors. Additional weighting factors can be applied, both during the run time and afterwards with separate post processing programs. We demonstrated our approach using ideal fluence detectors, called USRBDX detector, as a single pixel detector. The detector was located 1 cm from the object. USRBDX detector is an ideal boundary crossing detector that counts the photons crossing a boundary, their energy, and their

בס"ד

angle with respect to the normal to the surface of the detector. We set a 0.5·0.5 cm$^2$ USRBDX detector as a one-way fluence detector and considered the total fluence.

For the simulation of GI reconstruction with scattered photons we set up 4 additional detectors. The detectors were defined in a similar fashion to the direct USRBDX detector, but located at altitudes of 45°, 90°, 135° and 180° in respect to the optical axis. The detectors were numbered #1-#5, respectively, where detector #1 measures the radiation transmitted through the object. Figure 3 presents the schematic description of the imaging setup with all 5 detectors.

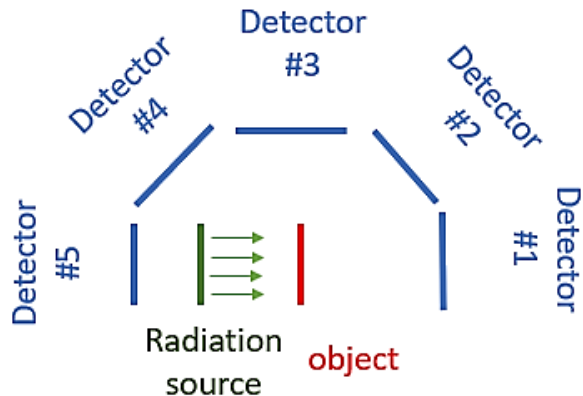

**Figure 3: Simulation setup for Ghost Imaging with scattered photons**

### *2.4 The GI reconstruction process*

The images were reconstructed based on 2,000 realizations, assuming 10 million initial primaries in each realization using Matlab 2021b and the "Total Variation optimization - an Alternating minimization Algorithm for Augmented Lagrangian functions" (TVAL3) library [8, 26, 27]. The TVAL3 parameters remained constant for all reconstructions.

To quantify the reconstruction quality we used the Peak Signal to Noise Ratio (PSNR) value. The equation for PSNR value is presented in Eq.1.

$$PSNR = 10 \cdot \log_{10} \frac{v_{max}^2 \cdot M \cdot N}{\sum_{M,N}[Ref(m,n) - Image(m,n)]^2} \quad (1)$$

where $v_{max}$ is the maximum pixel value in the images, M, N are the numbers of rows and columns in the images, Ref is the reference image, and Image is the reconstructed image. We set the reference image as a binary matrix of 50*50 pixels. In this matrix a value of 1 indicates the presence of the object while a 0 value represents the presence of the surrounding air, as shown in figure 4.

בס"ד

## 3 RESULTS

### *3.1 Transmission Ghost Imaging*

Figure 4 presents the reference image and the reconstruction of transmission GI. It is evident that the reconstruction is in excellent agreement with the reference.

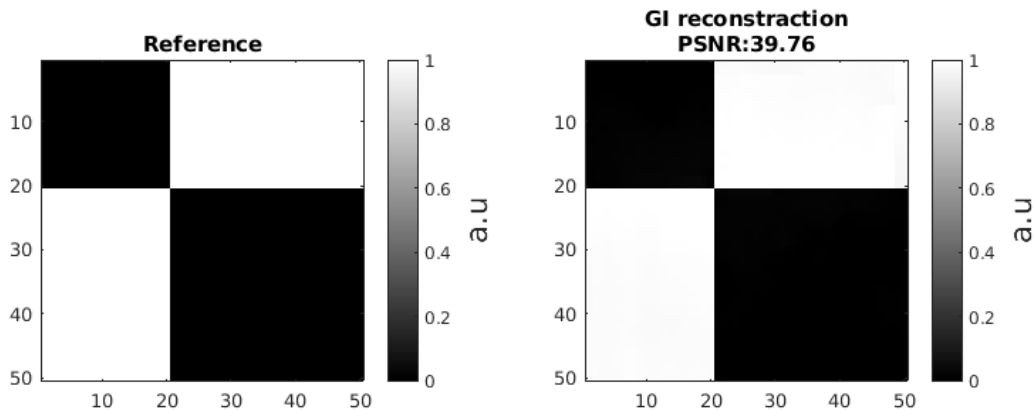

**Figure 4: Monte Carlo simulation of Ghost Imaging reconstruction**

To quantitatively assess the agreement between the reconstruction and the ground truth, we analyse the average PSNR value as a function of realizations used, as depicted in Figure 5. It can be observed that the PSNR value reaches its maximum after approximately 1300 realizations and then remain constant. These results align with previous experiments that utilized TVAL3 [15]. The compression ratio is defined as the number of pixels divided by the number of used realizations. In the example we explore in the present work we found that compression ratio is 2500/1300~2, which agrees with the values observed in the experiments [15].

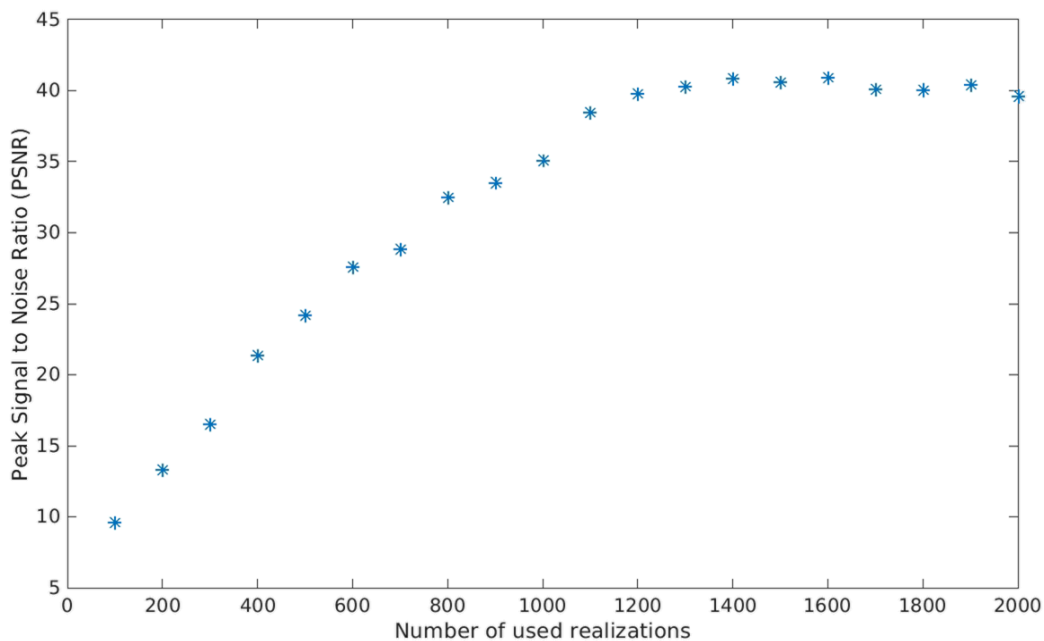

**Figure 5: Peak Signal to Noise Ratio (PSNR) value as a function of the number of used realizations of the Ghost Imaging reconstruction**

### 3.2 *Ghost Imaging with scattered photons*

Realizing that any signal proportional to the electron density of the object can in principle provide information about the structure of the object, we simulated GI reconstruction using scattered photons. These simulations must include all types of radiation matter interactions, highlighting the essential nature of our approach for accurately simulating the physical processes involved in scattered radiation-based GI.

To further illustrate advantages of our approach and to demonstrate its ability to serve a tool for designing systems beyond conventional transmission measurements, we conducted simulations of scattered GI using detectors #2-#5, positioned as described in Fig. 3. The results are summarized in Fig. 6, where we observe that the reconstructed images are the inverse of the image obtained through the direct transmission reconstruction in Fig.4. This result is anticipated because, in the case of scattered photons, the object itself– the scattering medium – serves as the source of the radiation detected by the detectors.

To assess the quality of the images reconstructed with the scattered radiation and compare them with the direct transmission reconstruction, we calculated the PSNR value with the inverse of the reconstruction, which we refer as the "inverse PSNR". From Fig. 6 it is evident that the reconstructions are in good agreement with the inverse direct imaging in figure 4. However, the inverse PSNR value of the scattered photon GI is lower, and the contrast is comparatively poor compared to the direct GI reconstruction. Therefore, our simulations indicate that further optimization is required to enhance the scattered photon GI reconstruction.

בס"ד

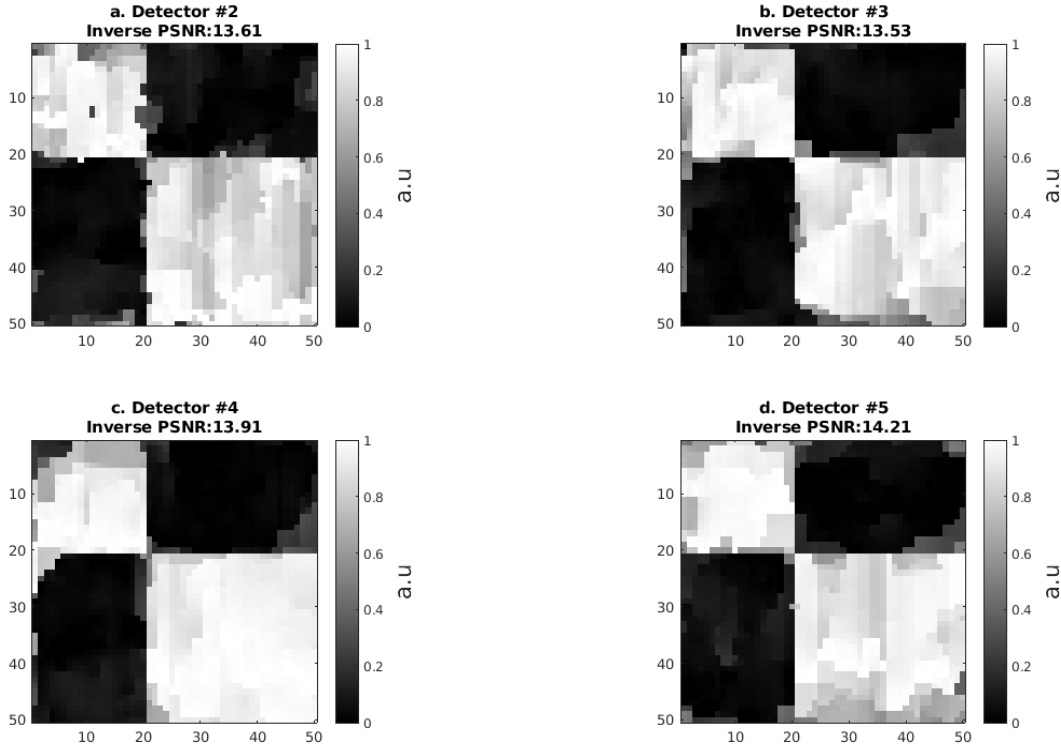

**Figure 6: Ghost Imaging reconstruction with scattered photons.** These results show the reconstructed images from the detector #2 - #5. As expected, the reconstructed images from scattered photons are the inverse of the image obtained by direct transmission reconstruction.

## 4 CONCLUSIONS AND FUTURE WORK

We have demonstrated benefits of MC simulations for the analysis and design of GI systems. Since traditional methods for GI simulations rely on oversimplified models that neglect important radiation-matter interactions such as scattering, they are unable to accurately simulate GI at very high photon energies, which are commonly employed in X-ray imaging techniques. In contrast, MC simulations for GI provide more comprehensive information about radiation-matter interactions that is lost with the oversimplified models. This information is essential for the design and analysis of GI systems at high-photon energies. Our approach can be extended to other imaging modalities and to GI with particle radiation such as protons or neutrons. Moreover, the MC simulations we have used here can be employed to non-imaging related calculations such as energy absorption and dose calculations, which are important for assessing radiation-induced damage in electronic components and ensuring radiation safety.

בס"ד

## 5 FUNDING

This research was supported by the Pazy Research Foundation.

## 6 DATA AVAILABILITY

The datasets generated during and/or analyzed during the current study are available from the corresponding author on reasonable request.

בס"ד

**8 AUTHOR CONTRIBUTIONS**
- Conceptualization: RHS, LE, SS
- Methodology: RHS, YK, OS, YF
- Investigation: RHS, YK, OS, YF
- Visualization: RHS
- Monte-Carlo simulations: RHS
- Supervision: SS
- Writing – original draft: RHS, SS
- Writing – review & editing: RHS, YK, OS, YF, LE, SS

**9 COMPETING INTERESTS**

Authors declare that they have no competing interests.